\documentclass[sigconf]{acmart}
\AtBeginDocument{%
  }

\copyrightyear{2026}
\acmYear{2026}
\setcopyright{usgovmixed}
\acmConference[WWW '26]{Proceedings of the ACM Web Conference 2026}{April 13--17, 2026}{Dubai, United Arab Emirates}
\acmBooktitle{Proceedings of the ACM Web Conference 2026 (WWW '26), April 13--17, 2026, Dubai, United Arab Emirates}
\acmPrice{}
\acmDOI{10.1145/3774904.3793034}
\acmISBN{979-8-4007-2307-0/2026/04}
\usepackage{subfigure}
\usepackage{amsmath}
\usepackage{amsthm}
\usepackage{amssymb}
\usepackage{bm}
\usepackage{booktabs}
\usepackage{multirow}

\usepackage{algorithm}
\usepackage{algpseudocode}
\usepackage{enumitem}
\setlist[itemize]{leftmargin=*}
\newcommand{\m}{\textsc{EnergyUCB}}

\begin{document}

\title{Online GPU Energy Optimization with Switching-Aware Bandits}

\author{Xiongxiao Xu}
\authornote{Work primarily performed while interning at Argonne National Laboratory.}
\affiliation{%
  \institution{Illinois Institute of Technology}
  \city{Chicago}
  \state{IL}
  \country{USA}
}
\email{xxu85@hawk.illinoistech.edu}

\author{Solomon Abera Bekele}
\affiliation{%
  \institution{Argonne National Laboratory}
  \city{Lemont}
  \state{IL}
  \country{USA}
}
\email{sbekele@anl.gov}

\author{Brice Videau}
\affiliation{%
  \institution{Argonne National Laboratory}
  \city{Lemont}
  \state{IL}
  \country{USA}
}
\email{bvideau@anl.gov}

\author{Kai Shu}
\affiliation{%
  \institution{Emory University}
  \city{Atlanta}
  \state{GA}
  \country{USA}
}
\email{kai.shu@emory.edu}

\renewcommand{\shortauthors}{Xiongxiao Xu et al.}

\begin{abstract}
Energy consumption has become a bottleneck for future computing architectures, from wearable devices to leadership-class supercomputers. Existing energy management techniques largely target CPUs, even though GPUs now dominate power draw in heterogeneous high performance computing (HPC) systems. Moreover, many prior methods rely on either purely offline or hybrid offline and online training, which is impractical and results in energy inefficiencies during data collection. In this paper, we introduce a practical online GPU energy optimization problem in a HPC scenarios. The problem is challenging because (1) GPU frequency scaling exhibits performance–energy trade-offs, (2) online control must balance exploration and exploitation, and (3) frequent frequency switching incurs non-trivial overhead and degrades quality of service (QoS). To address the challenges, we formulate online GPU energy optimization as a multi-armed bandit problem and propose {\m}, a lightweight UCB-based controller that dynamically adjusts GPU core frequency in real time to save energy. Specifically, {\m} (1) defines a reward that jointly captures energy and performance using a core-to-uncore utilization ratio as a proxy for GPU throughput, (2) employs optimistic initialization and UCB-style confidence bonuses to accelerate learning from scratch, and (3) incorporates a switching-aware UCB index and a QoS-constrained variant that enforce explicit slowdown budgets while discouraging unnecessary frequency oscillations. Extensive experiments on real-world workloads from the world's third fastest supercomputer Aurora show that {\m} achieves substantial energy savings with modest slowdown and that the QoS-constrained variant reliably respects user-specified performance budgets.
\end{abstract}





\maketitle

\section{Introduction}
Energy consumption has become a central challenge for digital society, with direct consequences for environmental sustainability, economic resilience, and global equity~\cite{berndt1990energy}. As the scale and ubiquity of computation continue to grow, from everyday handheld gadgets~\cite{hussein2022adaptive,sarmad2022reducing}, such as smartphones and wearable health devices, to the world’s most powerful supercomputers~\cite{atchley2023frontier}, such as the Aurora system at Argonne National Laboratory, the energy footprint of computing infrastructures increasingly affects national power grids as well as the accessibility and sustainability of digital services worldwide. For example, the Aurora supercomputer, the third-fastest system globally as of January 2026, reaches a peak power draw of about 60\,MW, comparable to the electricity demand of a mid-sized U.S.\ city\footnote{https://www.anl.gov/aurora}. Similarly, in 2022 the RIKEN Center was forced to power down one-third of the Fugaku supercomputer due to soaring energy prices in Japan\footnote{https://www.fujitsu.com/global/about/innovation/fugaku}. These examples highlight an urgent need for energy-aware computing across the stack from devices to data centers.

\begin{figure}[t]
    \centering
    \hspace{-0.3cm}
    \subfigure[]{
    	\begin{minipage}{0.24\textwidth}
   		 	\includegraphics[width=1\textwidth]{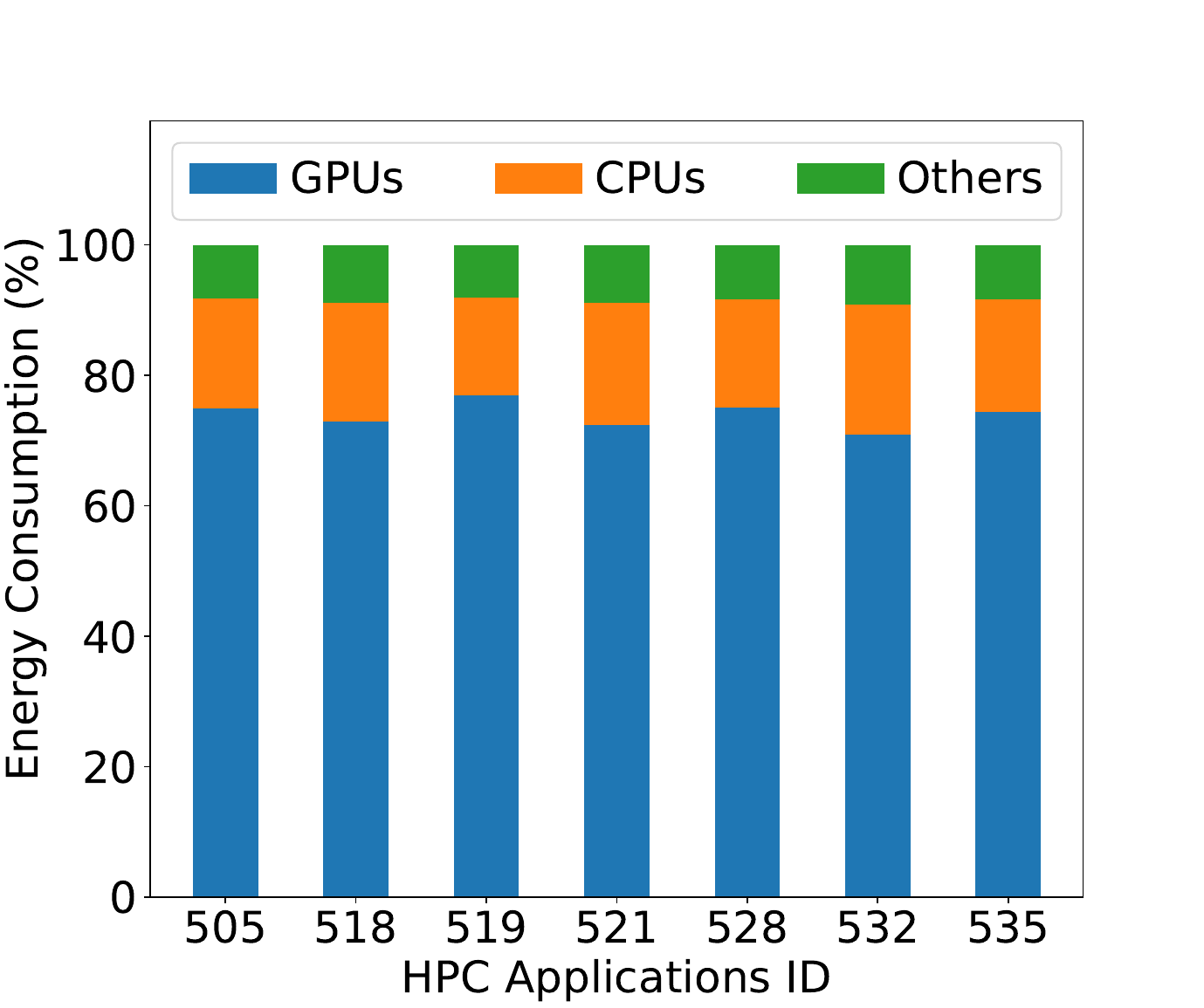}
    	\end{minipage}
            \label{fig:mot_energy_dis}
    }
    \hspace{-0.8cm}
    \subfigure[]{
    	\begin{minipage}{0.24\textwidth}
   		 	\includegraphics[width=1\textwidth]{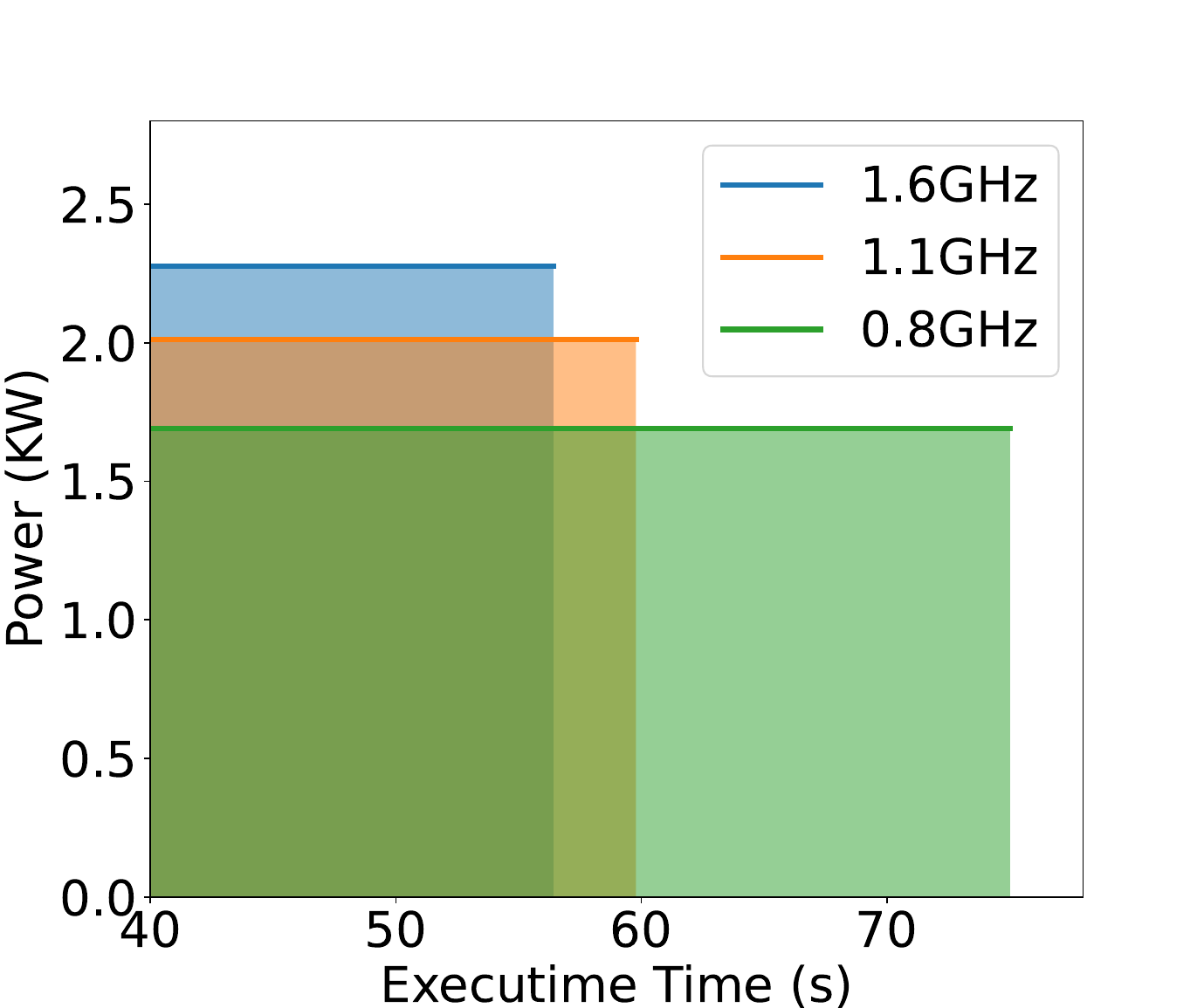}
    	\end{minipage}
            \label{fig:mot_perf_energy}
    }
    \vspace{-0.3cm}
    \caption{(a) The distribution of energy consumption of CPUs, GPUs, and other components for multiple HPC applications on a compute node from Aurora supercomputer. (b) Performance-energy trade-off in GPUs for the HPC application \texttt{pot3d}. At 1.6\,GHz, energy consumption is 128.46\,(kJ)=2.277\,(kW)$\times$56.42\,(s); at 1.1\,GHz, energy consumption is 120.21\,(kJ)=2.011\,(kW)$\times$59.78\,(s); at 0.8\,GHz, energy consumption is 126.78\,(kJ)=1.690\,(kW)$\times$75.02\,(s).}
    \vspace{-0.7cm}
\end{figure}

While progress has been made in reducing the energy consumption of computing systems, existing efforts have predominantly focused on CPUs~\cite{cerf2021sustaining,wang2021online}. However, the growing importance of GPUs, particularly in AI model training, such as for large language models (LLMs) that demand substantial GPU computational resources, has shifted the focus. As of July 2025, nine of the top ten fastest supercomputers in the TOP500 list are GPU-powered\footnote{https://top500.org}. In these heterogeneous computing systems, GPUs have become the dominant energy consumers. Figure~\ref{fig:mot_energy_dis} plots the energy consumption distribution across components of a compute node in the Aurora supercomputer during SPEChpc 2021 benchmark runs, including GPUs, CPUs, and other components (e.g., memory). The results highlight that GPUs consume significantly more energy than CPUs and other parts. For example, when running the SPEChpc application \texttt{pot3d}, GPUs account for 75.10\% of the total energy consumption, over four times that of CPUs, which consume only 16.55\%. This emphasizes the most importance of optimizing GPU energy usage for effective energy management in heterogeneous computing systems. Furthermore, existing solutions primarily operate in either purely offline settings or hybrid offline and online settings, which are often impractical and lead to excessive energy consumption. In real-world systems, the process of collecting prior data itself incurs additional energy overhead. To investigate these limitations, this paper introduces a novel online energy optimization problem. 

However, online energy optimization in GPUs presents several challenges. First, there is an inherent trade-off between performance and energy efficiency in processors. Modern processors employ various energy-saving techniques, with dynamic voltage and frequency scaling (DVFS) being one of the most widely adopted. DVFS adjusts a processor's frequency and corresponding voltage to achieve energy efficiency. Although reducing frequency lowers power consumption, it often leads to performance degradation, resulting in longer execution time. Since energy is the product of power and execution time, this creates a complex performance-energy trade-off across different frequencies. As illustrated in Figure~\ref{fig:mot_perf_energy}, when the GPU core frequency decreases from 1.6\,GHz to 1.1\,GHz, the GPU power drops from 2.277~kW to 2.011~kW, while the execution time for the HPC application \texttt{pot3d} increases from 56.42s to 59.78s. Consequently, energy consumption decreases from 128.46~kJ to 120.21~kJ. However, when the frequency is further reduced to 0.8\,GHz, power decreases to 1.690~kW, but the execution time significantly extends to 75.02s, causing energy consumption to rise from 120.21~kJ to 126.78~kJ.

Second, online optimization introduces an exploration \& exploitation dilemma across discrete frequency levels. At the beginning of a job, the controller has no prior information about the GPU's behavior at different frequencies. It must actively explore these options and rely on feedback from hardware counters (e.g., energy consumption and core utilization) to build an on-the-fly profile of performance and energy. This process creates a tension between exploration and exploitation: exploration focuses on testing frequencies that have rarely been used, while exploitation leverages observations from frequencies already tested. Because each trial consumes time and energy, it is crucial for the algorithm to strike an effective balance between exploration and exploitation.

Third, changing GPU frequency is not free. Each switch incurs latency and additional energy, and aggressive oscillations between nearby frequencies can degrade quality of service (QoS) for end users. An online controller must therefore reason about switching costs. For many latency-sensitive applications, it should respect explicit slowdown budgets specified by users. Designing a method that learns good frequency policies, limits unnecessary switching, and enforces QoS constraints is a key challenge for making GPU energy optimization practical in production HPC systems.

To address the aforementioned challenges, we formulate online GPU energy optimization as a multi-armed bandit problem and propose {\m}, a lightweight UCB-based controller that dynamically adjusts GPU core frequency in real time. In this framework, frequency options are modeled as arms, and feedback from GPU hardware counters serves as the reward. We adopt the ratio of core-to-uncore utilization as a real-time proxy for GPU performance and design a reward that balances energy consumption and execution progress within each time step. Building on the merits of the bandit framework, {\m} offers a principled solution to manage the exploration \& exploitation dilemma during online frequency control. 
Extensive experiments demonstrate that {\m} consistently reduces GPU energy consumption with modest slowdown, positioning it as a practical solution for large-scale computing systems. The main contributions of this paper are summarized as follows:
\begin{itemize}
    \item \textbf{Problem.} We introduce a new online GPU energy optimization task, and formalize it as a multi-armed bandit problem. 
    \item \textbf{Algorithm.} We develop {\m}, a lightweight UCB-based controller with optimistic initialization for fast bootstrapping and a switching-aware UCB index to reduce unnecessary frequency oscillations. We further present a QoS-constrained extension to enforce user-specified slowdown budgets.
    \item \textbf{Evaluation.} We evaluate {\m} across diverse workload running on Aurora supercomputer with two metrics \textit{saved energy} and \textit{energy regret}. The extensive results demonstrate that {\m} achieves substantial energy savings on Aurora while maintaining performance.
    \item \textbf{Social Impact.} An idealized full-scale deployment on a system like Aurora could potentially mitigate the daily energy footprint of 9{,}000 U.S. residents or 69{,}000 people (around 130~t CO$_2$/day) in under-resourced regions, highlighting the potential societal benefits of energy-efficient GPU computing at scale. This impact would be significantly higher if extended across the global population of GPU devices in commercial and research infrastructures.
\end{itemize}

\vspace{-0.3cm}
\section{Preliminaries}\label{sec:preli}
In this section, we introduce the architecture of the Intel PVC, basis of multi-armed bandits, and problem definition.
\subsection{The Aurora Node Architecture}
\begin{figure}[t]
    \centering
    \includegraphics[width=0.40\textwidth]{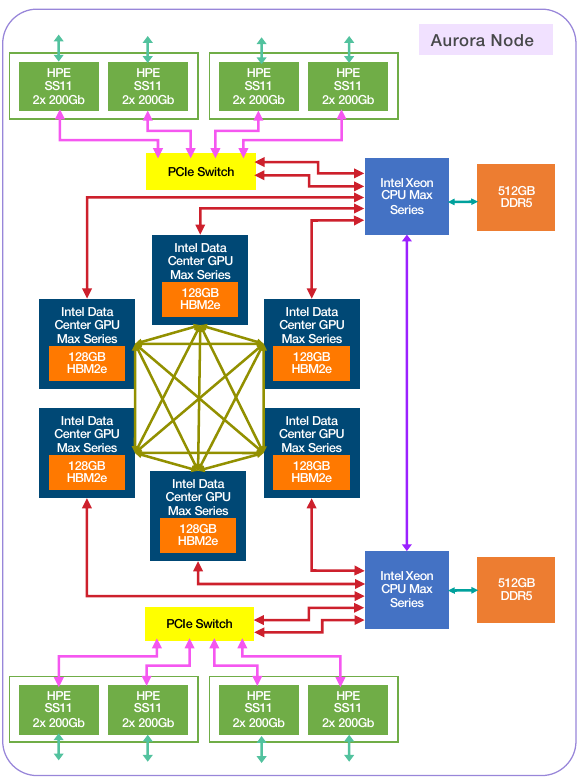}
    \caption{The architecture of an Aurora node.}
     \vspace{-0.5cm}
    \label{fig:PVC}
\end{figure}

Figure~\ref{fig:PVC} shows that a single Aurora node comprises of two Intel Xeon CPU Max Series processors, known as Sapphire Rapids or SPR, equipped with on-package High Bandwidth Memory (HBM), and six Intel Data Center GPU Max Series, also known as Ponte Vecchio or PVC. Each Xeon CPUs have 52 cores, with two hardware threads per core, and are outfitted with 64GB of HBM. The PVC is built on the Xe Core architecture. Each Xe core is composed of 8 vector and 8 matrix engines, supported by 512 KB of L1 cache. They are interconnected using the Intel XeLink interfaces. Every node includes 8 HPE Slingshot-11 Network Interface Cards (NICs). A group of 16 Xe cores forms a slice, and 4 such slices are combined with a substantial L2 cache and 4 HBM2E memory controllers to create a stack or tile. In this work, we use Aurora’s GPUs as a case study, but the framework is generic and can be applied to other GPU architectures.

\subsection{Multi-Armed Bandits}
\textbf{Basic Formulation}. In many real-world scenarios, it is important to balance the exploration \& exploitation dilemma, i.e., exploiting the current accumulated observations and exploring new knowledge through searching unknown spaces. A classic formulation of the decision-making framework to address the exploration \& exploitation dilemma is the $K$-armed bandit problem. Formally, there are $K$ finite arms. At each time step $t\in\{1,2,...,T\}$, one arm out of the $K$ arms is pulled, and let $I_t\in\{1,2...,K\}$ be the arm pulled at time step $t$. After $I_t$ is pulled, the associated reward $r_t$ of the arm $I_t$ is observed by the bandit algorithm. Given a fixed time cost $T$, the goal of the algorithm is to maximize the total rewards over a sequence of time steps $T$ as follows:
\begin{equation}\label{eq:reward}
    \max\sum_{t=1}^{T}r_t
\end{equation}
The decision $I_t$ at each time step $t$ involves choosing between exploiting the arm with the highest accumulated rewards until time $t-1$ and exploring other arms to gather more knowledge.
% about their potential rewards.
\\\textbf{Reward Model.}
The generated reward of each arm $i$ follows a probability distribution $D_i\in\{D_1, D_2,...,D_K\}$ with mean $\mu_i\in\{\mu_1,\mu_2,...,\mu_K\}$. When pulling an arm $i$, the reward will be sampled independently from the distribution $D_i$. In other words, given the history up to time $t-1$ and the choice of arm $I_t$ at time $t$, the reward is drawn independently with respect to the distribution of the chosen arm. In a formal way, let $H_{t-1} = \{(I_1,r_1,), (I_2,r_2,), ..., (I_{t-1},r_{t-1})\}$ denote the history of observations until time $t-1$. The expected reward for arm $i$ can be written as follows:
\begin{equation}
    E[r_t|H_{t-1}, I_t=i] = \mu_i
\end{equation}
It implies that the reward generated is randomly disturbed by noise.
\\\textbf{Cumulative Regret.}
The performance of the bandit algorithm is measured by the gap between the evaluated algorithm and the Oracle algorithm which can choose the best arm all the time. Formally, let $I^*=\arg\max_{i=1,2,...,K}\mu_i$ and $\mu^*=\mu_{I^*}$ be the
index of the best arm selected by Oracle algorithm and the associated highest expected reward. We define the cumulative regret at time $T$ as follows:
\begin{equation}\label{eq:regret}
    R(T)=\sum_{t=1}^{T}(\mu^*-\mu_{I_t})
\end{equation}
The goal of the bandit algorithm is to minimize regret in Eq.~\ref{eq:regret} or equally maximize reward in Eq.~\ref{eq:reward}. 

\subsection{Problem Definition}
Following the above notations, we give a formal problem definition for the online energy consumption in GPUs.
\begin{center}
\fbox{\parbox[c]{.95\linewidth}{\textbf{Online Energy Optimization in GPUs.}
Given an application running on GPUs at the default maximum frequency, the task is to dynamically adjust the GPU core frequency to minimize energy consumption while ensuring timely completion. At each time step $t$, the algorithm selects a frequency and observes data from hardware counters, leading to a reward $r_t$. By incorporating this feedback, the algorithm accumulates the knowledge and refines its frequency adjustment strategy. The process continues until the application completes at time $T$.
}}
\end{center}
There are two key points to emphasize. (1) The problem is set in a fully online environment. This means that the algorithm cannot access any prior information regarding profiles of GPUs and applications under frequencies. Instead, the algorithm must learn and adapt by directly interacting with real-time data from the GPUs' hardware counters. (2) The time cost $T$ varies across different applications and frequencies. Since each application requires a distinct workload, their completion times $T$ will differ. Additionally, the history of frequency changes affects the processing speed of GPUs, leading to variations in the completion time $T$.

\section{Methodology}
In this section, we formulate online GPU energy optimization as a multi-armed bandit problem and introduce {\m}, a lightweight bandit algorithm that dynamically adjusts GPU core frequencies real-time. 
{\m} is designed to (i) balance the performance–energy trade-off when selecting frequencies, (ii) resolve the exploration–exploitation dilemma without offline training, and (iii) explicitly account for the hardware cost of switching frequencies.

\subsection{Modeling GPU Online Energy Consumption}
We model online energy consumption in GPUs as a multi-armed bandit problem, including frequency modeling, reward formulation, and completion time modeling.
\\\textbf{Frequency Modeling.} Modern circuit technologies integrate voltage regulates in a chip, supporting DVFS. In this regard, GPUs in Aurora system support software controllable, discrete voltage and frequency states that can be adjusted to meet specific performance and energy goals. There are finite discrete GPU core frequencies available in the system. In a formal way, let $f_i$ be a frequency and $K$ be the number of frequencies. We can model multiple frequency choices $\{f_1,f_2,\dots,f_K\}$ as a set of arms $\{1,2,\dots,K\}$. For example, the GPU core frequencies can be adjusted from 0.8\,GHz to 1.6\,GHz with 0.1\,GHz interval, i.e., $f_i\in\{0.8\text{\,GHz},0.9\text{\,GHz},\dots,\text{1.6\,GHz}\}$. By modeling frequencies as arms, we define the exploration space of the algorithm as a finite set of $K$ frequency options. The bounded space enables the algorithm quickly identify the optimal frequency, thus ensuring energy savings. Note that we do not have to model the state, a concept that is required in reinforcement learning (RL)~\cite{kaelbling1996reinforcement}. The burdensome design of states in RL leads to long convergence time~\cite{beggs2005convergence}, during which a large quantity of energy will be wasted.
\\\textbf{Reward Formulation.}
The design of the reward function is pivotal in guiding the convergence direction of the algorithm. On the Aurora supercomputer, the default configuration operates at the maximum frequency. Our goal is to minimize energy consumption by dynamically adjusting the GPU core frequency. However, reducing the frequency can degrade performance, leading to an increase in execution time $T$ and potentially increasing overall energy consumption. This intricate trade-off between performance and energy necessitates a carefully crafted approach.

GPUs have a monotonic energy counter and a timestamp counter to track energy consumption at each time step $t$. Therefore, the energy consumption between two timestamps $(t_1, t_2)$ can be calculated by taking the difference between the respective records. However, explicit lightweight performance counters that indicate the progress of offload kernels are not available in the GPUs. To address the limitation, we propose to leverage the utilization metrics provided by the GPUs as a performance proxy. In detail, the GPUs have an active-time counter to record when the resource is actively running workloads~\cite{level-zero} between two timestamps $(t_1, t_2)$. The utilization is calculated by taking the percentage of active time between the two timestamps. It is essential in systems as it indicates the component currently used by the workload, allowing us to infer the behaviors of the workload.

To approximate the performance, we leverage the ratio of GPU core utilization (including compute engines) and GPU uncore utilization (including copy engines responsible for data movement) as a proxy. A higher ratio indicates that the workload is compute-bound and more sensitive to core frequency scaling, while a lower ratio suggests the workload is memory-bound and more sensitive to data movement. We formally define the reward $r_t$ at time $t$ as:
\begin{equation}
    r_t=-E_t*R_t=-E_t*\frac{{UC}_t}{UU_t}.
\end{equation}
Here, $E_t$, ${R}_t$, ${UC}_t$, and ${UU}_t$ denote energy consumption, core-to-uncore utilization, core utilization, and uncore utilization of a GPU within time $t$. 
\\\textbf{Completion Time.} 
In the classical bandit problem, the time horizon $T$ is a predefined constant to determines the stopping condition. However, in our context, $T$ is application-dependent and proportional to the application's execution time. Crucially, $T$ is influenced by frequency choices during execution: higher frequencies improve performance and reduce execution time, whereas lower frequencies degrade performance and increase execution time. Consequently, we define the stopping condition $T$ as the completion of the entire application, with $T$ dynamically determined by the application's characteristics and frequency selections. Let $\{p_1,p_2,...,p_k\}$ represent the application's progress at each frequency within a time step. When a frequency $f_i$ is selected, the remaining workload $S$ is computed based on the progress $p_i$ achieved under $f_i$. This process iteratively continues until the HPC workload is fully consumed.

\begin{algorithm}[t]
\caption{{\m}}
\label{alg:energyucb}
\begin{algorithmic}[1]
\Require Frequencies $\{f_1,\dots,f_K\}$; progress $\{p_1,\dots,p_K\}$; switching penalty $\lambda$; optimistic prior $\mu_{\mathrm{init}}$
\State Initialize $t \gets 1$, $S \gets 1$, $I_{\text{prev}} \gets 1$
\For{$i = 1$ to $K$}
    \State $n_{i,0} \gets 0$, $\hat{\mu}_{i,0} \gets \mu_{\mathrm{init}}$
\EndFor
\While{$S > 0$}
    \For{$i = 1$ to $K$}
        \State $\mathrm{SA\mbox{-}UCB}_{i,t} \gets \hat{\mu}_{i,t} + \alpha \sqrt{\dfrac{\ln t}{\max(1,n_{i,t})}} - \lambda \cdot \mathbf{1}\{ i \neq I_{\text{prev}} \}$
    \EndFor
    \State $I_t \gets \arg\max_i \mathrm{SA\mbox{-}UCB}_{i,t}$
    \State Set frequency to $f_{I_t}$; observe reward $r_t$ and progress $p_{I_t}$
    \State $n_{I_t,t} \gets n_{I_t,t} + 1$
    \State $\hat{\mu}_{I_t,t} \gets \hat{\mu}_{I_t,t} + \dfrac{r_t - \hat{\mu}_{I_t,t}}{n_{I_t,t}}$
    \State $S \gets S - p_{I_t}$; $I_{\text{prev}} \gets I_t$; $t \gets t + 1$
\EndWhile
\end{algorithmic}
\end{algorithm}

\subsection{{\m} Algorithm}
In online energy optimization, each reckless trial can lead to extended execution time and increased energy consumption. Since our problem is fully online and {\m} has no prior knowledge of applications, it is crucial to carefully design the initialization strategy, balance exploration and exploitation, and account for the non-trivial hardware cost of changing GPU core frequencies. 

Algorithm~\ref{alg:energyucb} summarizes the proposed {\m}. Initially, the algorithm performs an \emph{optimistic initialization} (lines 2--4) to encourage adaptive exploration. In the subsequent \emph{exploration \& exploitation} phase (lines 5--14), it uses a switching-aware index (defined below) to adaptively select frequencies until the application completes. This design allows {\m} to quickly identify near-optimal frequencies while keeping frequency transitions small.

\noindent\textbf{Exploration vs.\ exploitation.}
Pure exploration strategies (e.g., round-robin over all frequencies) can characterize the application under each setting but incur substantial energy loss during exploration. Pure exploitation strategies that always choose the empirically best frequency may miss better options due to insufficient observations of rarely chosen arms, leading to excessive energy usage. To address this dilemma, {\m} adopts an upper confidence bound (UCB) strategy, which provides a principled way to trade off exploration and exploitation without assuming prior knowledge of the reward distribution. {\m} maintains an empirical mean reward $\hat{\mu}_{i,t}$ and a pull count $n_{i,t}$ for each frequency $f_i$. The standard UCB index is
\[
\mathrm{UCB}_{i,t} = \hat{\mu}_{i,t}
+ \alpha \sqrt{\frac{\ln t}{\max(1,n_{i,t})}},
\]
where the first term encourages exploitation of high-reward frequencies, while the second term encourages exploration of less frequently selected ones.

\noindent\textbf{Optimistic initialization.}
{\m} adopts an optimistic initialization to encourage adaptive exploration under noisy measurement conditions. In large HPC systems, factors such as clock synchronization, temperature fluctuations, and network congestion~\cite{libri2016evaluation,acun2016variation,xu2024surrogate} can cause GPU hardware counters (e.g., energy counters) to report unstable values at early time steps, leading to high-variance observed rewards. In this setting, a naive warm-up that blindly tests each frequency once can be both noisy and wasteful, because every trial is executed on the real application. Instead of explicitly cycling through every frequency, {\m} initializes each frequency $f_i$ with an optimistic prior $\hat{\mu}_{i,0} = \mu_{\mathrm{init}}$, which makes all arms initially attractive under the UCB index above. As rewards are observed, these estimates are updated online and suboptimal frequencies are naturally phased out, allowing {\m} to accumulate information adaptively rather than through a fixed round-robin schedule.

\noindent\textbf{Switching-aware UCB.}
On Aurora, changing GPU core frequencies through the GEOPM runtime
interface incurs a measurable overhead in both energy and time. While a single switch is inexpensive relative to the total application energy, frequent toggling between nearby frequencies can accumulate unnecessary overhead and extend execution time, making such behavior undesirable in a production setting. To capture this practical constraint, {\m} employs a switching-aware UCB rule.

Let $I_{t-1}$ denote the frequency index used at the previous time step. We define the switch-aware index for arm $i$ at time $t$ as:
\begin{equation}
    \mathrm{SA\mbox{-}UCB}_{i,t}
    = \hat{\mu}_{i,t} + \alpha \sqrt{\frac{\ln t}{\max(1,n_{i,t})}} - \lambda \,\mathbf{1}\{ i \neq I_{t-1} \},
    \label{eq:saucb}
\end{equation}
where $\lambda \ge 0$ is a switching penalty and $\mathbf{1}\{\cdot\}$ is the
indicator function. The decision rule in the exploration-exploitation phase
then selects
\begin{equation}
    I_t = \arg\max_{i \in \{1,\dots,K\}} \mathrm{SA\mbox{-}UCB}_{i,t}.
\end{equation}
When $\lambda = 0$, {\m} reduces to the standard UCB policy. For
$\lambda > 0$, the new frequency guarantees a sufficiently large
expected gain to compensate for the penalty, thus discouraging
unnecessary transitions and reflecting the hardware switching cost.

\subsection{Constrained {\m} for QoS Guarantee}\label{sec:energyucb_qos}
Quality of service (QoS) is crucial in our setting, since minimizing energy inevitably increases execution time to some extent. Some users or applications may require an explicit performance guarantee, such as minimizing energy subject to at most 10\% slowdown. To support such requirements,
we extend {\m} to a constrained variant that operates only on frequencies whose estimated slowdown is within a user-specified budget $\delta$.

Let $f_{\max}$ denote the maximum GPU frequency and let $\hat{p}_i$ be the estimated progress per decision interval under frequency $f_i$. We define
the relative slowdown of arm $i$ as
$s_i = 1 - \hat{p}_i / \hat{p}_{\max}$. Given a performance budget
$\delta \in [0,1)$, we construct the feasible set
\[
    \mathcal{K}_\delta = \{ i \mid s_i \le \delta \}.
\]
Constrained {\m} runs the policy only over arms
in $\mathcal{K}_\delta$, i.e., it selects
$I_t = \arg\max_{i \in \mathcal{K}_\delta} \mathrm{SA\mbox{-}UCB}_{i,t}$. This extension allows users to specify an interpretable constraint like $\delta = 0.05$ (at most 5\% slowdown), while {\m} automatically searches for the energy-efficient configuration within the budget.

\begin{table*}[!t]
    \centering
    \caption{Energy consumption (Unit: kJ) results on various HPC applications. Best results are shown in bold.}
    \begin{tabular}{c|c|c|c|c|c|c|c|c|c}
    \hline
        \textbf{Methods} & \textbf{lbm} &\textbf{tealeaf} &\textbf{clvleaf} &\textbf{miniswp} &\textbf{pot3d} &\textbf{sph\_exa} &\textbf{weather} &\textbf{llama} &\textbf{diffusion}\\ \hline
        \textbf{1.6\,GHz} & 93.94  & 109.79 & 100.65 & 187.13 & 131.13 & 1,353.41 & 134.61 & 1,277.71 & 772.21 \\
        \textbf{1.5\,GHz} &  \textbf{93.71} & 107.09 & 98.72 & 177.10 & 129.11 & 1,259.65 & 128.43 & 1,257.58 & 771.50 \\
        \textbf{1.4\,GHz} & 97.42 & 105.52 & 94.72 & 171.60 & 127.24 & 1,216.60 & 125.52 & 1,211.42 & 770.91 \\
        \textbf{1.3\,GHz} & 99.88 & 105.37 & 91.61 & 167.25 & 125.75 & 1,191.01 & 122.80 & 1,294.05 & 766.59 \\
        \textbf{1.2\,GHz} & 104.42 & 101.65 & 90.99 & 164.45 & 126.66 & 1,163.51 & 121.75 & 1,177.68 & 771.07 \\
        \textbf{1.1\,GHz} & 109.59 & 99.81 & 90.35 & 161.72 & \textbf{123.38} & 1,146.37 & \textbf{120.47} & 1,202.81 & 751.82 \\
        \textbf{1.0\,GHz} & 116.04 & \textbf{98.61} & \textbf{88.41} & 160.17 & 125.19 & 1,116.52 & 122.52 & \textbf{1,114.29} & 766.73 \\
        \textbf{0.9\,GHz} &124.28 & 99.10 & 89.00 & 160.15 & 125.45 & 1,107.28 & 123.38 & 1,360.93 & 805.50 \\
        \textbf{0.8\,GHz} &131.61 & 100.59 & 91.23 & \textbf{158.74} & 128.79 & \textbf{1,090.24} & 122.97 & 1,210.13 & \textbf{747.20} \\
        \hline
        \textbf{RRFreq} &105.76 & 103.24 & 93.24 & 168.22 & 129.12 & 1,187.86 & 125.07 & 1,282.21 & 781.75 \\
        \textbf{$\epsilon$-greedy} & 100.86 & 100.88 & 91.32 & 168.28 & 130.08 & 1,106.65 & 123.24 & 1,273.75 & 785.02 \\
        \textbf{EnergyTS} & 99.17 & 100.79 & 91.76 & 168.02 & 129.50 & 1,104.55 & 123.95 & 1,268.31 & 784.18 \\ \hline
        \textbf{RL-Power} & 99.42 & 102.11 & 92.85 & 170.08 & 130.94 & 1,132.27 & 124.92 & 1,248.66 & 778.94 \\
        \textbf{DRLCap} & 101.88 & 103.97 & 93.77 & 175.92 & 131.86 & 1,168.33 & 125.41 & 1,231.56 & 785.53\\
        \textbf{DRLCap-Online} & 108.95 & 108.04 & 96.23 & 181.27 & 135.62 & 1,243.73 & 128.89 & 1,261.81 & 796.15\\
        \textbf{DRLCap-Cross} & 98.85 & 102.84 & 92.02 & 169.80 & 134.94 & 1,183.86 & 126.35 & 1,291.55 & 789.25\\ \hline
        \textbf{{\m}} & \textbf{94.25} & \textbf{99.06} & \textbf{90.08} & \textbf{162.72} & \textbf{124.93} & \textbf{1,095.89} & \textbf{122.73} & \textbf{1127.17} & \textbf{750.90} \\ 
        \textbf{Saved Energy} & -0.31 & 10.73 & 10.57 & 24.41 & 6.2 & 257.52 & 11.88 & 150.54 & 21.31\\
        \textbf{Energy Regret} & 0.54 & 0.45 & 1.67 & 3.98 & 1.55 & 5.65 & 2.26 & 12.88 & 3.7 \\
    \hline
    \end{tabular}\label{tab:main results}
\end{table*}

\section{Experiments}
In this section, we introduce the details of experiments, including experimental setup and experimental results. The code is here\footnote{https://github.com/XiongxiaoXu/EnergyUCB-Bandit}.
\subsection{Experimental Setup}
We provide details of experimental platform, dataset collection processes, baselines and evaluation metrics.
\\\textbf{Experimental Platform.} We conduct experiments on a node of the Aurora system shown in Figure~\ref{fig:PVC} with GEOPM (Global Extensible Open Power Manager)~\cite{geopm2017} for telemetry monitoring and frequency control. GEOPM is a versatile tool that allows users to monitor system energy and power consumption while optimizing hardware settings to achieve energy or performance objectives. GEOPM consists of two primary components: the GEOPM Service and the GEOPM Runtime. The GEOPM Service provides user-level access to detailed hardware metrics and control options through a secure interface. Concurrently, the GEOPM Runtime leverages the GEOPM Service to adjust hardware settings based on real-time hardware metrics and feedback from various applications profiling.
\\\textbf{Dataset Collection.} The SPEChpc 2021 benchmark suite~\cite{spechpc2021} is evaluated in the Aurora supercomputer. We specifically employ the MPI+OMP target offloading version of the tiny benchmarks to fully leverage all six GPUs in the system. The suite consists of seven benchmarks: \texttt{505.lbm}, \texttt{518.tealeaf}, \texttt{519.clvleaf}, \texttt{521.miniswp}, \texttt{528.pot3d}, \texttt{532.sph\_exa}, and \texttt{535.weather}. Additionally, we run two representative LLMs and diffusion model in the Aurora system: \texttt{Llama-2}~\cite{touvron2023llama} and \texttt{Stable Diffusion XL}~\cite{podell2023sdxl}. We set a 10ms sampling period for monitoring. On each application, we test all available frequencies of GPUs and collect the corresponding traces.
\\\textbf{Baselines.}
To the best of our knowledge, this is the first work to address the problem of online GPU energy consumption optimization without relying on any offline training. To this end, we compare {\m} with the below three types of baselines:
\begin{itemize}
    \item \textbf{Static Algorithms:} \textit{\{1.6\,GHz, 1.5\,GHz,..., 0.8\,GHz\}} represent the available frequency options for GPU cores on the Aurora supercomputer where the maximum frequency 1.6\,GHz is the default setting. Each frequency setting is static, meaning that GPU cores maintain this frequency throughout the entire execution time.
    \item \textbf{Dynamic Algorithms:} \textit{RRFreq (Round-Robin Frequency)} cycles through each frequency in a circular order at each time $t$. \textit{$\epsilon$-greedy} explores less frequently chosen options with probability $\epsilon$ and exploits the frequencies that have the highest reward with probability $1-\epsilon$. \textit{EnergyTS (Thompson Sampling)} is a Bayesian-based approach to balance exploration \& exploitation by maintaining continuously updating a posterior distribution over the reward of each GPU frequency. 
    \item \textbf{Reinforcement Learning Algorithms:} \textit{RL-Power}~\cite{wang2021online} is an online RL-based power management method originally designed for CPU power capping. We retain its learning and decision mechanism to our setting while restricting the action space to GPU core frequencies and constructing the state from GPU hardware counters. \textit{DRLCap}~\cite{wang2024drlcap} is a deep RL framework for GPU frequency capping that combines offline pre-training with online adaptation to optimize energy efficiency across architectures. To adapt \textit{DRLCap} to our workloads, we allocate the first 20\% of each execution for training and deploy the learned policy on the remaining 80\%. For a fair comparison with fully online methods, the energy consumed during the remaining 80\% of execution is scaled by a factor of 1.25. We also evaluate two \textit{DRLCap} variants: \textit{DRLCap-Online}, which learns purely online on the target benchmark, and \textit{DRLCap-Cross}, which is pre-trained on other benchmark suites and evaluated on the target benchmark.
\end{itemize}
\textbf{Evaluation Metrics.} 
We employ two metrics in the evaluation.
\begin{itemize}
    \item \textit{Saved Energy} measures the reduction in energy consumption achieved by {\m} relative to the default maximum frequency (1.6\,GHz). This metric reflects the practical benefit of deploying {\m} on real systems.
    \item \textit{Energy Regret} captures how close {\m} is to the best possible static configuration. It is defined as the difference between the energy consumed by {\m} and the minimum energy among all static frequencies. In an online setting, some exploration is unavoidable, so no dynamic algorithm can exactly achieve this minimum, i.e., energy regret must be larger than 0.
\end{itemize}
\textbf{Implementation Details.} The available frequency options are \{0.8\text{\,GHz}, 0.9\text{\,GHz}, \dots, 1.6\text{\,GHz}\}, totaling $K = 9$ choices. The frequency adjustment interval is set to 10 ms, matching the sampling period of GEOPM. We repeat experiments 10 times and take the average values to mitigate randomness.

\begin{figure}[t]
    \centering
    \hspace{-0.3cm}
    \subfigure[\texttt{tealeaf} application]{
    	\begin{minipage}{0.24\textwidth}
   		 	\includegraphics[width=1\textwidth]{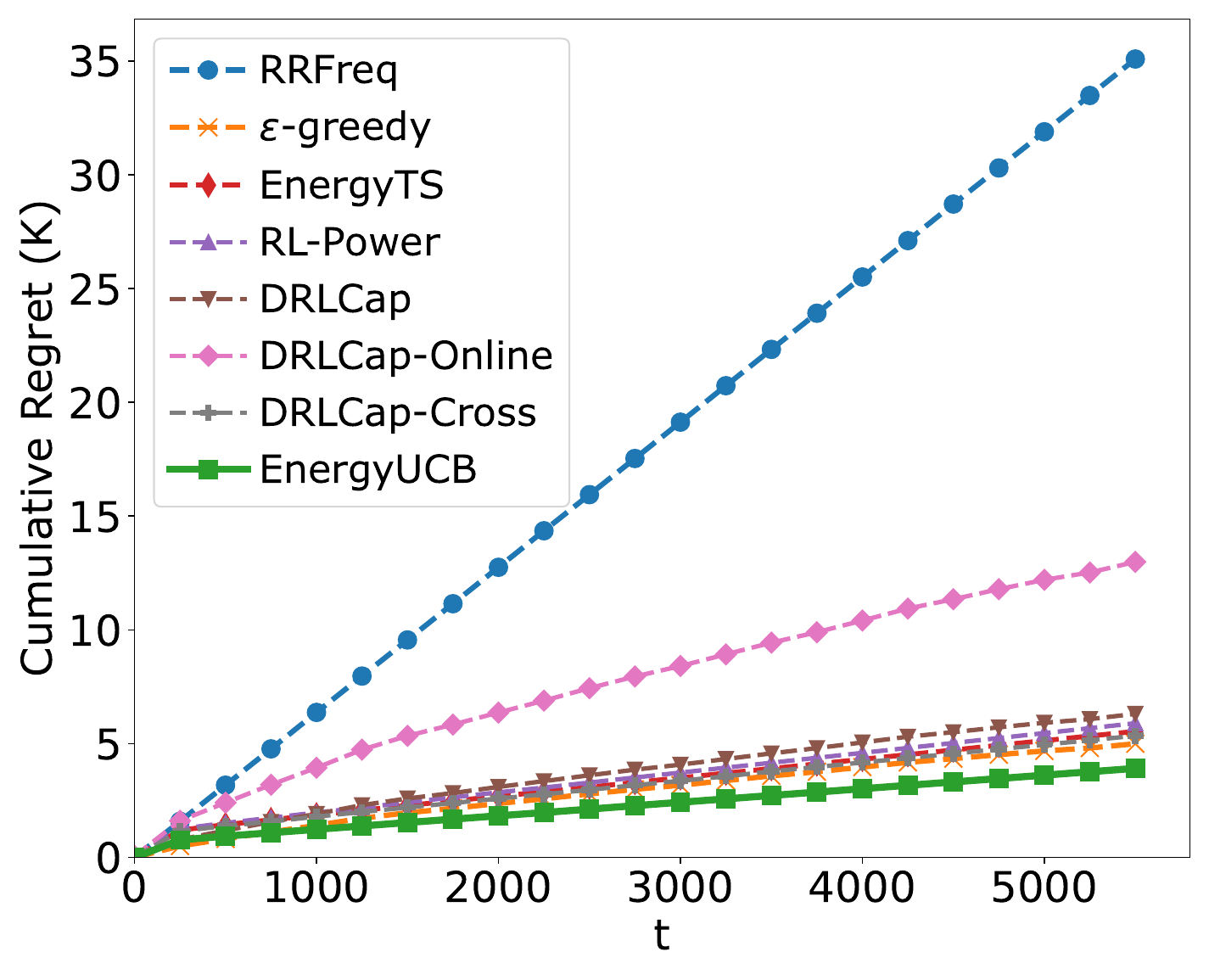}
    	\end{minipage}
    }
    \hspace{-0.35cm}
    \subfigure[\texttt{miniswp} application]{
    	\begin{minipage}{0.24\textwidth}
   		 	\includegraphics[width=1\textwidth]{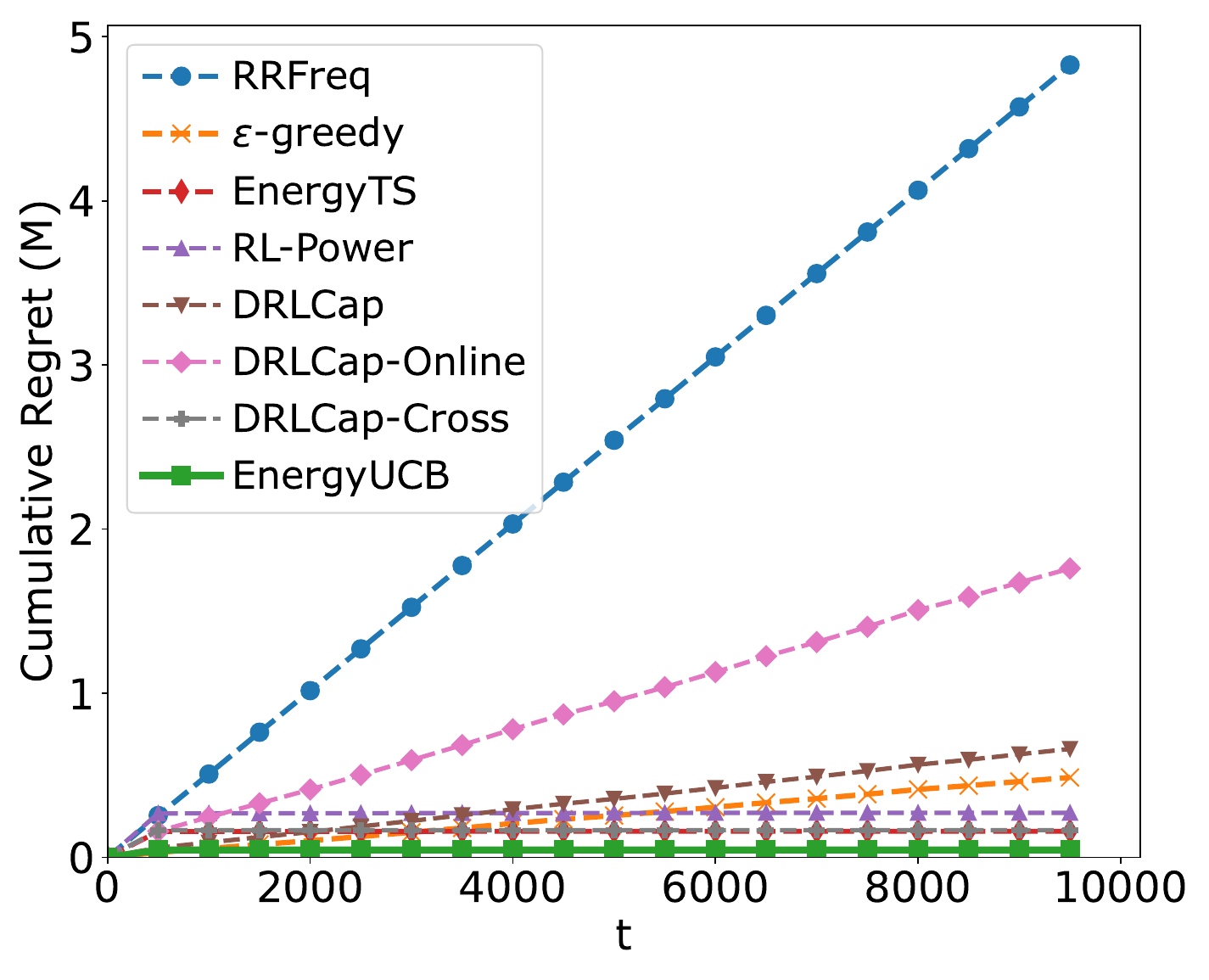}
    	\end{minipage}
    }
    \vspace{-0.3cm}
    \caption{Comparison of cumulative regret.}
    \label{fig:regret}
    \vspace{-0.5cm}
\end{figure}

\subsection{Experimental Results}
In this subsection, we analyze experimental results, including energy consumption and cumulative regret analysis.
\\\textbf{Comparison of Energy Consumption.}
We compare {\m} with the baselines w.r.t. the energy consumption. As shown in Table~\ref{tab:main results}, we have the following key findings:
\begin{itemize}
\item \textbf{There is no single optimal static frequency for all HPC applications.} For example, GPUs achieve the lowest energy consumption at 1.5\,GHz for \texttt{lbm}, whereas for \texttt{miniswp} and \texttt{sph\_exa}, the optimal frequency is 0.8\,GHz. This variation arises because different applications, which can be classified as either compute-bound or memory-bound~\cite{wang2021online}, respond differently to frequency scaling. Specifically, \texttt{lbm} is compute-bound, so lowering its frequency significantly increases execution time, leading to higher energy consumption. As a result, its optimal frequency is near the maximum. In contrast, \texttt{miniswp} is memory-bound, meaning that frequency reduction has little impact on execution time but lowers energy consumption, making its optimal frequency closer to the minimum.

\item \textbf{{\m} achieves significant energy savings compared to the default maximum frequency setting at Aurora.} 
Across nearly all applications, {\m} reduces energy compared to the default 1.6\,GHz at Aurora. For instance, {\m} saves 257.52~kJ on a single node for \texttt{sph\_exa}. Scaling to 10,620 Aurora nodes, running \texttt{sph\_exa} would save enough electricity per day to support 9,149 U.S. residents or 69,342 people in underdeveloped regions~\cite{business2020world}.  
For LLM workloads \texttt{llama}, {\m} reduces the energy by 150.54~kJ per node, highlighting the high energy footprint of foundation-model inference and the necessity of energy-aware GPU scheduling for emerging AI applications.

\item \textbf{{\m} consistently achieves lower energy than dynamic and RL algorithms.}
Compared with \textit{EnergyTS}, {\m} reduces energy by 141.14~kJ on \texttt{llama} and 8.66~kJ on \texttt{sph\_exa}.  
Against RL methods, {\m} outperforms \texttt{DRLCap} by 104.39~kJ on \texttt{llama} and by 13.2~kJ on \texttt{miniswp}. For example, \texttt{DRLCap} consumes 1,231.56~kJ on \texttt{llama}, whereas {\m} uses only 1,127.17~kJ. These results show that {\m} resolves the exploration–exploitation dilemma more efficiently and adapts to workload characteristics faster than DRL-based baselines, which require long convergence to stabilize.

\item \textbf{The small energy regret indicates that {\m} is close to the “optimal algorithm”.} The average energy regret and the average minimum energy across all benchmarks are 3.63 and 403.89, respectively, yielding an energy regret of only 0.89\%. This demonstrates that {\m} closely approximates the best static configuration, even though limited exploration is unavoidable in an online setting.
\end{itemize}
\textbf{Cumulative Regret.}
We evaluate the cumulative regret of algorithms to assess their learning efficiency and convergence. As shown in Figure~\ref{fig:regret}, the regret of {\m} grows slowly and flattens out quickly, whereas the regrets of \textit{RRFreq} increase almost linearly over
time, with other dynamic and RL algorithms lying in between. This indicates that {\m} rapidly identifies and exploits the optimal frequency, while baseline methods keep wasting energy on suboptimal choices. For example, for \texttt{tealeaf}, when the time step $t$ reaches 4,000 (about 40 seconds), the cumulative regret of {\m} is only 1.99k, whereas \textit{RRFreq} incurs 25.51k. 

\vspace{-0.4cm}
\subsection{Ablation Study}
To assess the contribution of each component in {\m}, we compare it against two variants:
\textit{w/o Opt.~Ini.} (removing optimistic initialization) and
\textit{w/o Penalty} (removing the switching-aware penalty).
Table~\ref{tab:ablation_study} reports the results on the three most energy-intensive applications.
{\m} consistently achieves the lowest total energy consumption, confirming the effectiveness of both optimistic initialization and the switching-aware UCB.

\begin{table}[t]
    \centering
    \scalebox{0.9}{
    \begin{tabular}{lccc}
        \toprule
        & \textit{{\m} (kJ)} & \textit{w/o Opt. Ini. (kJ)} & \textit{w/o Penalty (kJ)} \\
        \midrule
        \textbf{sph\_exa} & 1,095.89$\pm$0.52  & 1,116.71$\pm$2.28 & 1,102.70$\pm$0.75 \ \\
        \textbf{Llama} & 1,127.17$\pm$0.76 & 1,199.18$\pm$1.83 & 1,133.42$\pm$0.32 \\
        \textbf{Diffusion} & 750.90$\pm$0.58 & 788.33$\pm$2.65 & 753.66$\pm$0.14 \\
        \bottomrule
    \end{tabular}
    }
    \caption{Ablation study of {\m}. 
    }
    \vspace{-0.8cm}
    \label{tab:ablation_study}
\end{table}

Removing optimistic initialization significantly increases energy consumption because the controller lacks reliable estimates in the early stage and performs excessive exploratory probes. 
For example, on \textsc{Llama}, the absence of optimistic initialization increases energy consumption by 72.01~kJ. Removing the switching-aware penalty also increases energy consumption (by 6.25~kJ on \textsc{Llama}), although the magnitude is smaller.
This difference is expected. The optimistic initialization component directly contributes to learning the optimal frequency faster, whereas the switching-aware penalty primarily discourages oscillations between adjacent frequencies, thus reducing energy and time overhead of {\m} itself. We analyze this
switching cost more in the following subsection.

\begin{figure}[b!]
    \centering
    \includegraphics[width=0.48\textwidth]{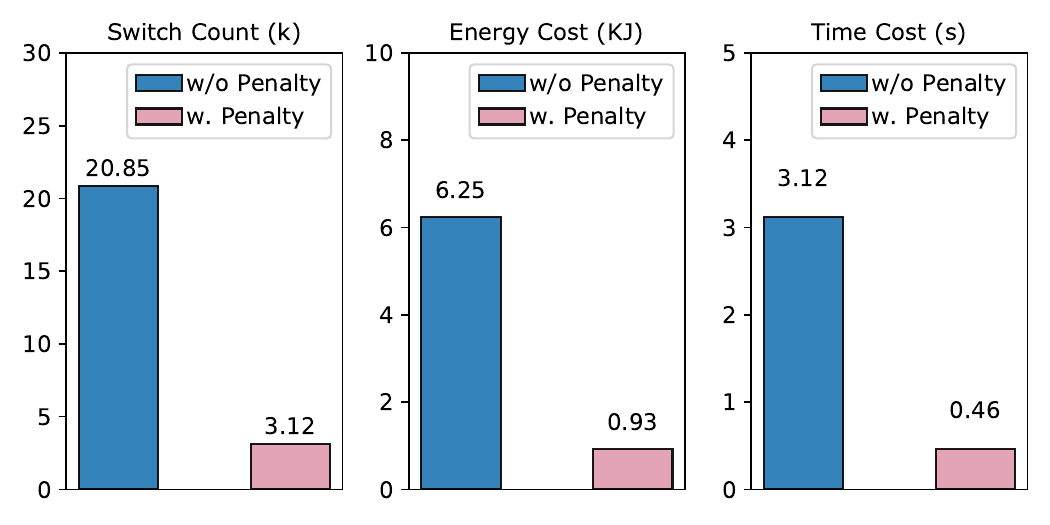}
     \caption{Switching cost analysis of {\m}.}
    \label{fig:switch_cost}
\end{figure}

\subsection{Switching Cost Analysis}
A practical concern is if the act of switching frequencies in {\m} introduces non-trivial time and energy overhead. In our implementation, GPU core
frequencies are adjusted through the GEOPM runtime interface, and each switch incurs approximately $150\,\mu\text{s}$ latency and 0.3~J per switch. Although a single switch is inexpensive, the cumulative cost can become substantial if a controller oscillates frequently across thousands of decision intervals.

Figure~\ref{fig:switch_cost} compares \textit{w/o Penalty} and \textit{with Penalty} on the \textsc{Llama} application. The switching-aware design reduces the number of frequency
changes from 20.85k to 3.12k, a 6.7$\times$ reduction. Consequently, the energy overhead introduced by {\m} decreases from 6.25~kJ to 0.93~kJ, and the additional execution time
induced by switching falls from 3.12~s to 0.46~s. These results confirm that while the per-switch overhead is small, the accumulated switching cost is non-negligible without explicit regularization. The switching-aware penalty in {\m} effectively suppresses unnecessary transitions, keeping the switching overhead tiny in practice while preserving the same near-optimal energy savings.

\begin{figure}[t]
    \centering
    \hspace{-0.3cm}    
    \subfigure[Reward Formulation]{
    	\begin{minipage}{0.24\textwidth}
   		 	\includegraphics[width=1\textwidth]{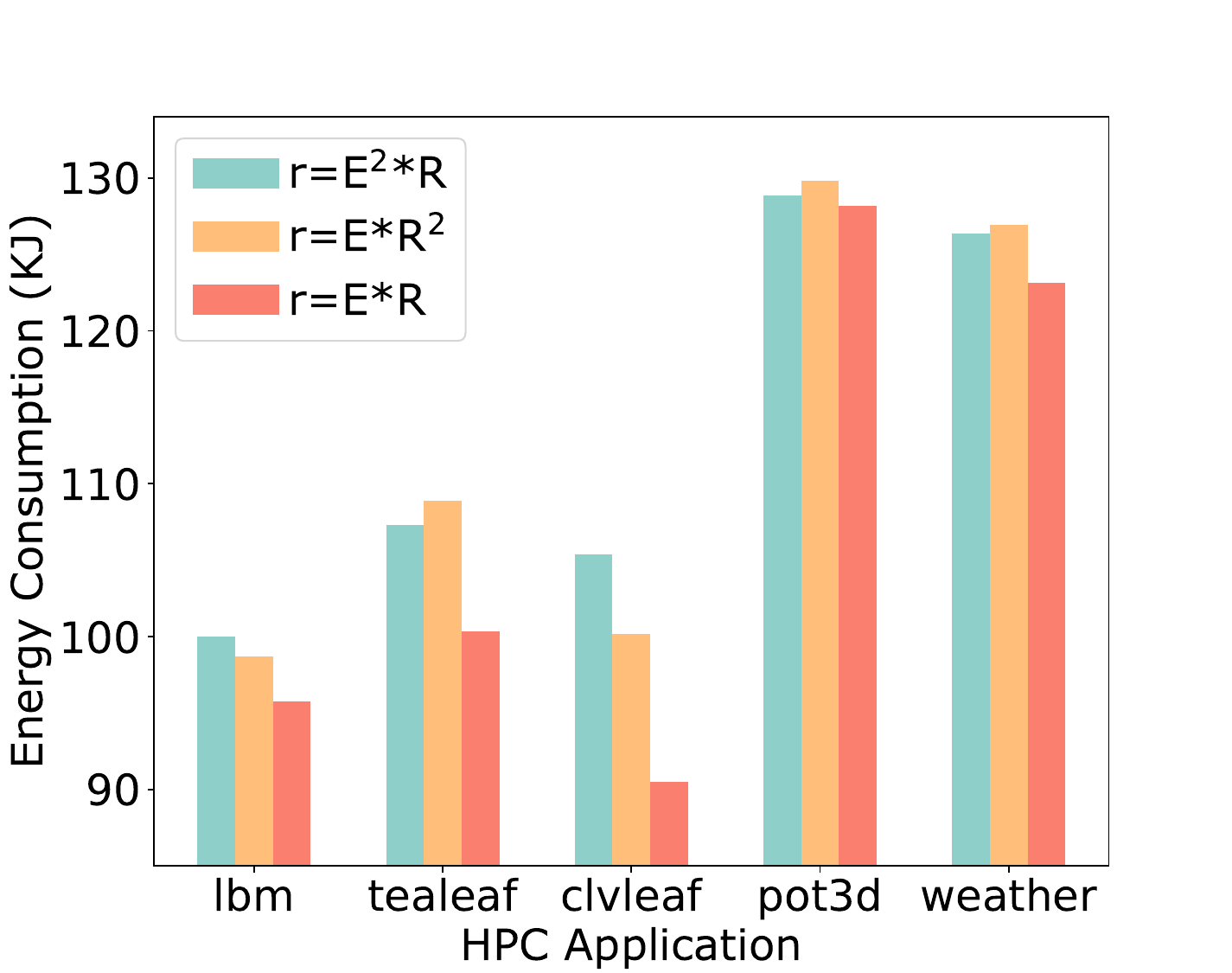}
    	\end{minipage}
    \label{fig:impact_reward}
    }
    \hspace{-0.5cm}
    \subfigure[QoS analysis]{
    	\begin{minipage}{0.24\textwidth}
   		 	\includegraphics[width=1\textwidth]{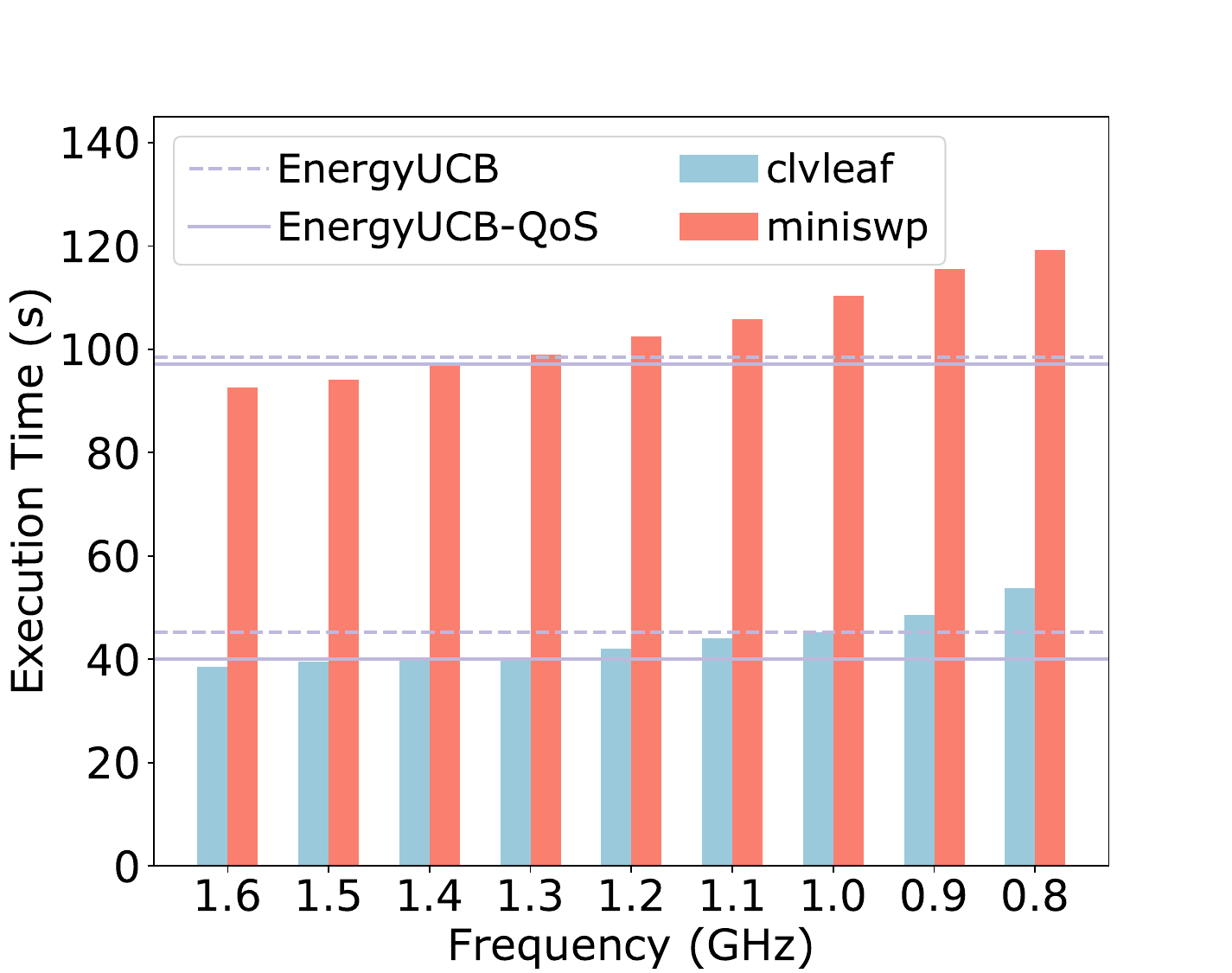}
    	\end{minipage}
    \label{fig:execution_time_b}
    }
    \vspace{-0.3cm}
    \caption{Impact of two reward functions and QoS analysis.}
    \vspace{-0.7cm}
\end{figure}

\subsection{Reward Formulation Analysis}
Our reward is designed as the product of energy and performance, $r = E * R$, to explicitly balance the energy–performance trade-off. In principle, different exponents can shift this balance. For instance, $r = E^{2} * R$ places more weight on energy reduction, whereas $r = E * R^{2}$ favors faster completion of HPC applications. We evaluate these three variants in Figure~\ref{fig:impact_reward}. Across all benchmarks, $E^{2} * R$ and $E * R^{2}$ yield higher energy than $E * R$. For example, on \texttt{miniswp}, $E^{2} * R$ and $E * R^{2}$ consume around 185~kJ, whereas $E * R$ uses about 167~kJ (an increase of 10.77\%). Similarly, on \texttt{clvleaf}, they require over 100~kJ compared to about 90~kJ with $E * R$ (an increase of 11.11\%). The squared terms amplify fluctuations in the noisy GPU environment, increasing variance in the observed rewards and slowing convergence. In contrast, the linear form $E * R$ yields more stable learning and the lowest overall energy usage.

\subsection{Quality of Service Analysis}
Aurora is configured by default to operate at the maximum GPU frequency (1.6\,GHz) to minimize execution time. Although unconstrained {\m} reduces energy consumption by dynamically lowering the frequency, it may introduce a modest performance penalty. Figure~\ref{fig:execution_time_b} reports execution time across static frequencies for two representative HPC applications and overlays the execution time achieved by both the unconstrained {\m} (dashed line) and the constrained QoS variant (solid line).

For both \texttt{clvleaf} and \texttt{miniswp}, unconstrained {\m} yields execution times comparable to running statically at 1.2–1.3\,GHz. This corresponds to only moderate slowdowns of 14.46\% and 6.26\%, respectively, relative to the 1.6\,GHz default. These results indicate that even without an explicit QoS requirement, {\m} naturally operates in a regime with limited performance degradation while achieving substantial energy savings.

We further evaluate the constrained variant introduced in Section~\ref{sec:energyucb_qos}. Under a slowdown budget of $\delta = 0.05$, constrained {\m} selects higher frequencies than the unconstrained version and consistently maintains execution time within the user-specified limit: 4.05\% slowdown on \texttt{clvleaf} and 4.82\% on \texttt{miniswp}. Importantly, constrained {\m} satisfies the QoS requirement without reverting to the maximum frequency; instead, it automatically identifies an operating point that respects the budget while still reducing energy consumption relative to the 1.6\,GHz baseline.

\vspace{-0.2cm}
\section{Related Work}\label{sec:related}
This work is related to two lines: (1) energy consumption optimization in CPUs/GPUs and (2) multi-armed bandits and its applications.

\subsection{Energy Optimization in CPUs/GPUs}
Energy optimization on CPUs has been widely studied~\cite{abera2018performance,bekele2019ml,wu2023performance}. Early work~\cite{zhu2003scheduling} proposes DVFS scheduling for energy-efficient multiprocessor systems.  
\cite{yang2015adaptive} leverages regression models to characterize performance–energy trade-offs. \cite{wang2021online} applies RL for runtime power control on CPUs. \cite{wu2023utilizing} ensembles linear, nonlinear, tree, and rule-based models to predict power consumption of CANDLE workloads. Compared with CPUs, GPU energy optimization is less explored~\cite{wang2010power} in the HPC environments.  
\cite{huang2019gpu} uses an offline neural model for GPU energy configuration based on static task characteristics. \cite{wang2021dynamic} proposes GPOEO to dynamically optimize GPU settings but requires offline training before deployment. Recently, RL-based GPU DVFS methods have been explored.  
DRLCap~\cite{wang2024drlcap} uses deep RL by combining offline and online learning for GPU frequency capping. 
DSO~\cite{wang2024dso} fuses static code analysis and dynamic information for GPU energy optimization. 
While effective, these methods rely on offline pretraining before deployment. 

\subsection{Multi-Armed Bandit and Its Applications}
Multi-armed bandit (MAB)~\cite{lattimore2020bandit} is a sequential decision-making framework that balances exploration and exploitation. It has been applied in diverse domains including clinical trials~\cite{durand2018contextual}, dynamic pricing~\cite{misra2019dynamic}, recommender systems~\cite{xu2021generalized}, anomaly detection~\cite{ding2019interactive}, and telecommunication~\cite{soemers2018adapting}. For example, in clinical trials~\cite{durand2018contextual}, bandit algorithms are used to dynamically adjust the allocation of treatments to patients based on observed outcomes, with the goal of optimizing patient welfare and efficiently identifying the most effective treatments. Some noteworthy variants consider additional factors, such as side features of arms, and conversational feedback. For instance, contextual bandits~\cite{li2010contextual} incorporate side information (context) to make treatment/policy decisions and have been widely used in personalized recommendations. Conversational bandits~\cite{zhang2020conversational} learn user preferences interactively by explicitly querying the user. 
Neural bandits~\cite{ban2021multi} leverage non-linear neural networks to approximate reward functions. 
However, no prior work study an online switching-cost aware bandit framework for GPU energy optimization in HPC environments.

\vspace{-0.2cm}
\section{Conclusion and Future Work}
In this paper, we initiate a new problem of online GPU energy consumption optimization. To address the problem, we formulate it as a multi-armed bandit framework and propose a lightweight UCB-based algorithm {\m} that effectively balances energy efficiency and performance. The extensive experiments across real-world workload from Aurora supercomputer demonstrates superior performance of the {\m} and meaningful energy-saving implications. 
One exciting future direction is considering practical physical factors, such as the cooling system of supercomputers.

\begin{acks}
This research utilized resources of the Argonne Leadership Computing Facility, a U.S. Department of Energy (DOE) Office of Science user facility at Argonne National Laboratory and supported by the U.S. DOE Office of Science-Advanced Scientific Computing Research Program, under Contract No. DE-AC02-06CH11357.
\end{acks}

\bibliographystyle{ACM-Reference-Format}
\bibliography{sample-base}

\end{document}